\documentclass[%
reprint,
superscriptaddress,
 amsmath,
 amssymb,
 aip,jcp,
]{revtex4-2}

\usepackage{mathrsfs}
\usepackage{bm}

\usepackage{subfigure}
\usepackage{graphicx}
\usepackage{dcolumn}
\usepackage{xr}
\usepackage{color}
\usepackage{braket}
\usepackage[version=4]{mhchem}
\usepackage[normalem]{ulem}

\usepackage{xr-hyper}

\newcommand{\hB}{\hat{B}}

\newcommand{\hS}{\hat{S}}

\begin{document}

\title{Spin/Phonon Dynamics in Single Molecular Magnets: I. quantum embedding.}

\author{
Nosheen Younas$^{1,2,3}$,
Yu Zhang$^{2}$,
Andrei Piryatinski$^{2}$,
Eric R. Bittner$^{1,3}$
 }

\address{
$^{1}$
Department of Physics, University of Houston, Houston, Texas 77204, United States\\
$^{2}$Theoretical Division, Los Alamos National Laboratory, Los Alamos, New Mexico 87545, United States\\
$^{3}$Center for Nonlinear Studies, Los Alamos National Laboratory, Los Alamos, New Mexico 87545, United States}

%
%
%

\begin{abstract}
Single molecular magnets (SMMs) and metal organic frameworks (MOFs) have attracted significant interest because of their potential in quantum information processing, scalable quantum computing, and extended lifetimes and coherence times. The limiting factor in these systems is often the spin dephasing caused by interactions and couplings with the vibrational motions of the molecular framework. 
This work introduces a systematic projection/embedding scheme to analyse spin-phonon dynamics in molecular magnets. 
This scheme consolidates all spin/phonon couplings into a few collective degrees of freedom. In this first of two papers, we establish the projection scheme and show its equivalence to the complete set of modes within second-order Redfield dynamics.
Using parameters obtained from ab initio methods for spin/phonon coupling via the Zeeman interaction, we apply this approach to compute the electronic spin relaxation times for a single molecule qubit \ce{VOPc(OH)8}, which features a single unpaired electron located on the central vanadium atom.
However, our general embedding scheme can be applied to any single-molecule magnet or qubit MOF with any coupling/interaction Hamiltonian.
This development offers a crucial tool for simulating spin relaxation in complex environments with significantly reduced computational complexity.
\end{abstract}


\maketitle

\section{Introduction}
Recent efforts have explored various physical systems for qubit architectures in the pursuit of practical quantum computing realisation, including superconducting circuits~\cite{Krantz2019}, trapped ions~\cite{Blatt:2012ux}, quantum dots~\cite{Loss1998}, single molecular magnets (SMMs)~\cite{Bertaina:2008wt}, and neutral atoms~\cite{Lukin:2001uo}. 
Recently, molecular magnets and metal-organic frameworks (MOFs) have attracted attention for their potential in quantum information processing, high-density magnetic data storage, and as magnetic resonance contrast agents~\cite{GaitaArino2019, wasielewski2020exploiting,Iqbal:2024aa, Yu:2020uo, Yamabayashi:2018uh, Jellen:2020tx}. 
These materials allow for the tailoring of magnetic behaviour through the synthetic versatility of molecular compounds. 
However, they face challenges because they are susceptible to molecular vibrations, which can affect qubit performance.
The interaction between spins and molecular vibrations plays a crucial role in molecular imaging, optoelectronics, and quantum technology, influencing applications such as MRI, energy harvesting, and understanding qubit decoherence~\cite{wasielewski2020exploiting}. This spin-phonon coupling is particularly significant in SMMs, known for their magnetic bistability and memory retention at low temperatures. MOF qubit engineering aims to slow the spin relaxation dynamics to enhance magnetic memory and quantum coherence.

\begin{figure}[!h]
 \subfigure[]{\includegraphics[width=0.45\columnwidth]{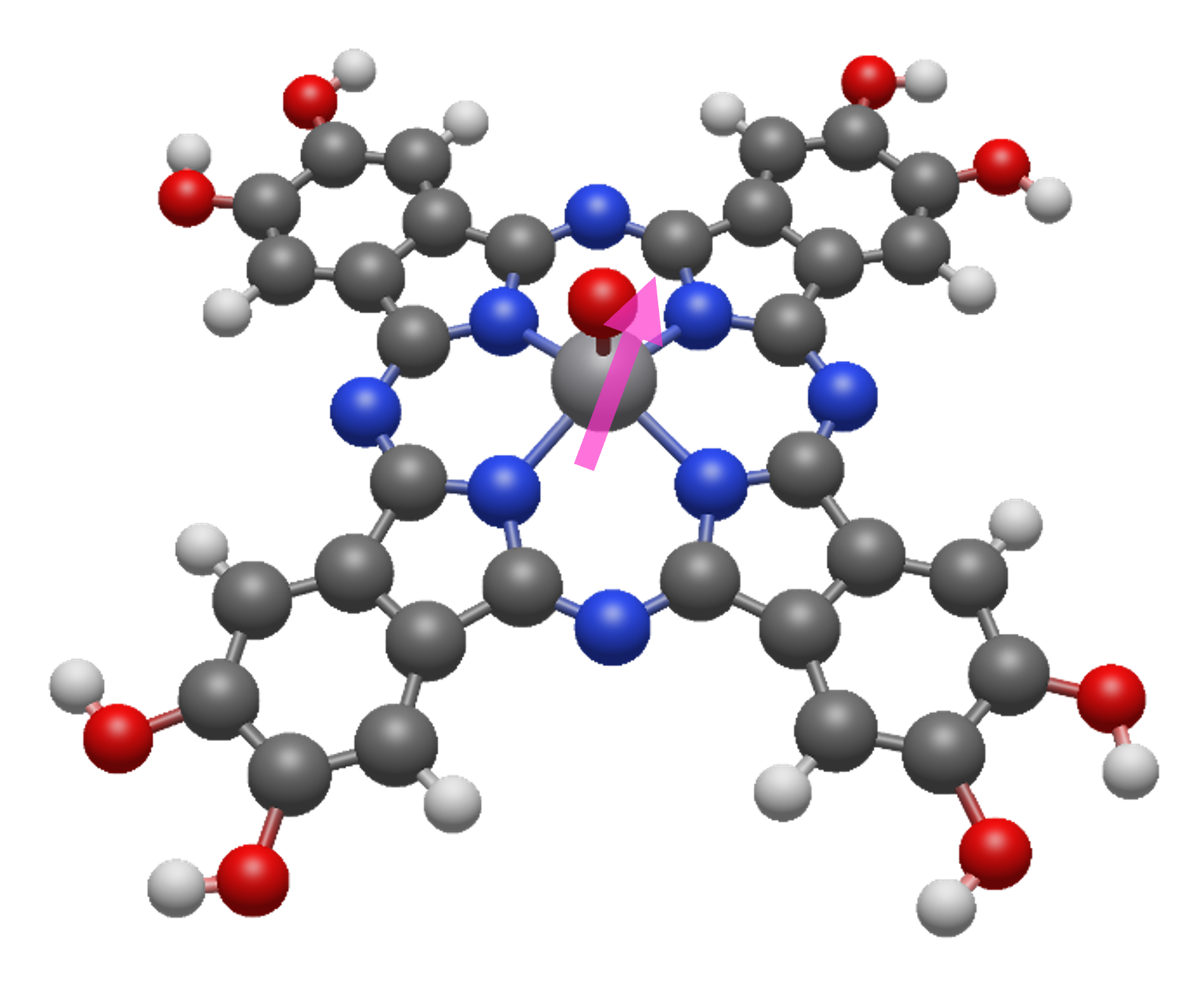}
    }
\subfigure[]{\includegraphics[width=0.45\columnwidth]{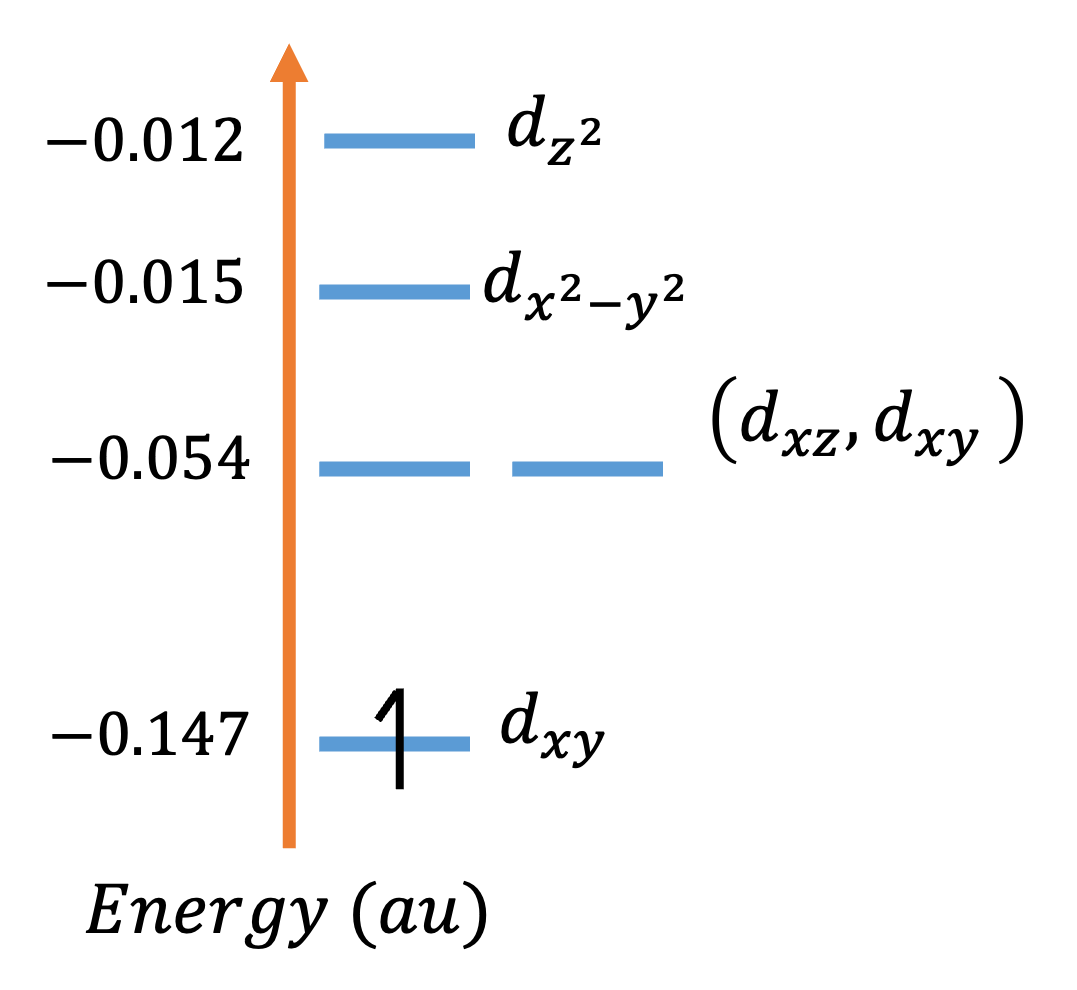}}
    \caption{(left) Single Molecule qubit of \ce{VOPc(OH)8} (a VOPc derivative) with carbon, nitrogen, hydrogen, and oxygen atoms shown in grey, blue, white, and red, respectively. The spin-qubit electron is shown with a pink arrow.
    (right) Energy level diagram for the d-orbitals for V$^{+4}$. Oxygen with a short double bond in vanadium IV causes $d_{xy}$ to be the lowest energy. Hence, due to $d^1$ configuration of V$^{+4}$, the spin-half qubit electron resides in $d_{xy}$ orbital.~\cite{Ryan2021jacs, Atzori:2016aa}}
    \label{fig_sph}
\end{figure}

Among various MOF architectures, vanadyl phthalocyanine (VOPc) is a prominent candidate for room-temperature quantum computing~\cite{Atzori:2016aa, Bonizzoni:2017aa, Malavolti:2018aa, Aziz:2011aa}. Fig.~\ref{fig_sph} shows a single VOPc molecule along with the d-orbital structure of the \ce{V^{+4}} centre hosting a single electron (in $d^1$ configuration) that is responsible for the qubit nature of this molecule.
First-principles calculations, especially density functional theory (DFT), have become essential in studying the electronic and magnetic properties of these molecular magnets~\cite{ja061798aa, Timco:2009tc, Chilton:2013aa, Chilton2015,Reta:2021aa}. These methods facilitate the investigation of spin-phonon dynamics without phenomenological parameters, aiding in the design of materials with tailored properties. However, challenges remain in accurately modelling these processes due to the high computational cost and complexity of spin-phonon interactions, which involve extensive degrees of freedom.

Highlighting these shortcomings underscores the value of exploring other first-principles approaches in the theoretical and computational literature that focus on phonon-assisted spin relaxation modelling. Lunghi \textit{et al.} recently used a
first-principles approach to model the
spin dynamics of \(\ce{VO(acac)2}\), focussing on first-order spin-phonon couplings through phonon-assisted modulation of Zeeman and dipolar Hamiltonians.\cite{Lunghi2017}  This
work validates using a low-order Taylor expansion approach for Redfield dynamics. However, even when modelling the complete crystal environment and phonons from the entire Brillouin zone using \(3\times 3\times 3\) supercells containing 1620 atoms at the DFT level, the estimates of spin-relaxation times remain orders of magnitude different from experimental results \cite{sciadv2019}. Similarly, other first-principles studies that incorporate Raman and Orbach mechanisms for phonon-assisted spin relaxation in single-molecule magnets (and qubits) only capture experimental relaxation values within a few orders of magnitude \cite{Lungi-direc_and_raman-VOacac, Lunghi-single-ion-magnets, Lunghi2017}. 
Despite quantitative disagreements, it is crucial to recognise that these first-principles studies successfully capture the qualitative aspects of phonon-assisted spin relaxation without relying on adjustable parameters. This provides a powerful tool for understanding quantum spin degradation mechanisms as well as engineering molecular architectures that protect these quantum properties.

Beyond first-principles efforts, various phenomenological methods have been employed to deduce the functional form of the \(T_1\) variation as a function of magnetic field and temperature by fitting curves to experimental data ~\cite{atzori-t1, Santanni:2021ts}.
Additionally, researchers have used ligand field methods to systematically analyse the impact of each phonon on spin relaxation through both direct and Raman processes. These analyses aim to uncover the role of each phonon based on phonon symmetry. However, these approaches require parameter adjustment to match the experimental values of \(T_1\) quantitatively \cite{Ryan2021jacs}.
Another notable effort has been presented by Aruachan \textit{et al.} using a semi-empirical approach by constructing a parameterised Redfield quantum master equation to describe the interaction of molecular spin qubits with lattice phonons and electron/nuclear spin baths. Based on the Haken-Strobl theory~\cite{Capek1985}, this approach treats system-reservoir interactions as stochastic fluctuations of the system Hamiltonian, with spin-lattice interaction and spin-spin interaction modelled by a fluctuating gyromagnetic tensor and a fluctuating local magnetic field, respectively. The master equation is then derived up to the first order in fluctuations. This semi-empirical procedure introduces fitting parameters into the bath spectral densities of the Redfield tensors, which must be calibrated by fitting to experimental measurements.~\cite{Aruachan}
Although this methodology yields excellent quantitative and qualitative agreement with the experimental data of $T_1$ and $T_2$ dynamics, it depends on prior experimental results for calibration.

Our approach in this paper follows the steps of other first-principles studies and extends them by isolating a handful of collective motions responsible for the spin-phonon interaction. Our first-principles model presented in this work is utilised in \textit{direct} spin-phonon interaction mechanism. 
Although this interaction is crucial at high fields and low temperatures (\(B>5\ T\) and \(T<20\ K\)), it does not prevail at low magnetic fields and intermediate temperatures. Consequently, deviations arise when comparing our results with the experimental data (Fig.~\ref{fig:redfield_combined}). However, such deviations are expected in first-principles calculations, which do not employ adjustable parameters.
We then introduce a quantum embedding approach using Singular Value Decomposition (SVD) to simulate spin-qubit dynamics efficiently. This method is inspired by mode projection schemes of Bittner 
\textit{et al.}~\cite{modeprojection2009jcp, Yang:2017aa,Yang:2018ab}. It reduces the number of phonon modes required for simulating the spin dynamics in molecular frameworks. Here we demonstrate the proficiency of this technique by simulating the spin lifetime dynamics in \ce{VOPc(OH)8} as an example. 
This approach also bridges the gap between SVD-based quantum embedding techniques and mode projection, creating an isomorphism between the two methodologies.

The following section
details the theoretical framework for spin-phonon interaction in molecular qubits and the development of the SVD-based mode projection scheme.  We also present details of the implementation, simulation results, and comparisons with previous studies. Finally, we summarises our findings and discusses future directions.
In our subsequent paper (Ref.~\cite{paper2})
we use the results presented herein to perform numerically exact quantum
dynamical calculations of the spin/phonon
system and show how these projected modes act as irreversible conduits of quantum
information from the spin into the thermal environment.

\section{Theory}\label{sec:theory}

\subsection{Spin-Vibrational Hamiltonian}
\begin{figure}
    \centering
    \includegraphics[width=0.5\textwidth]{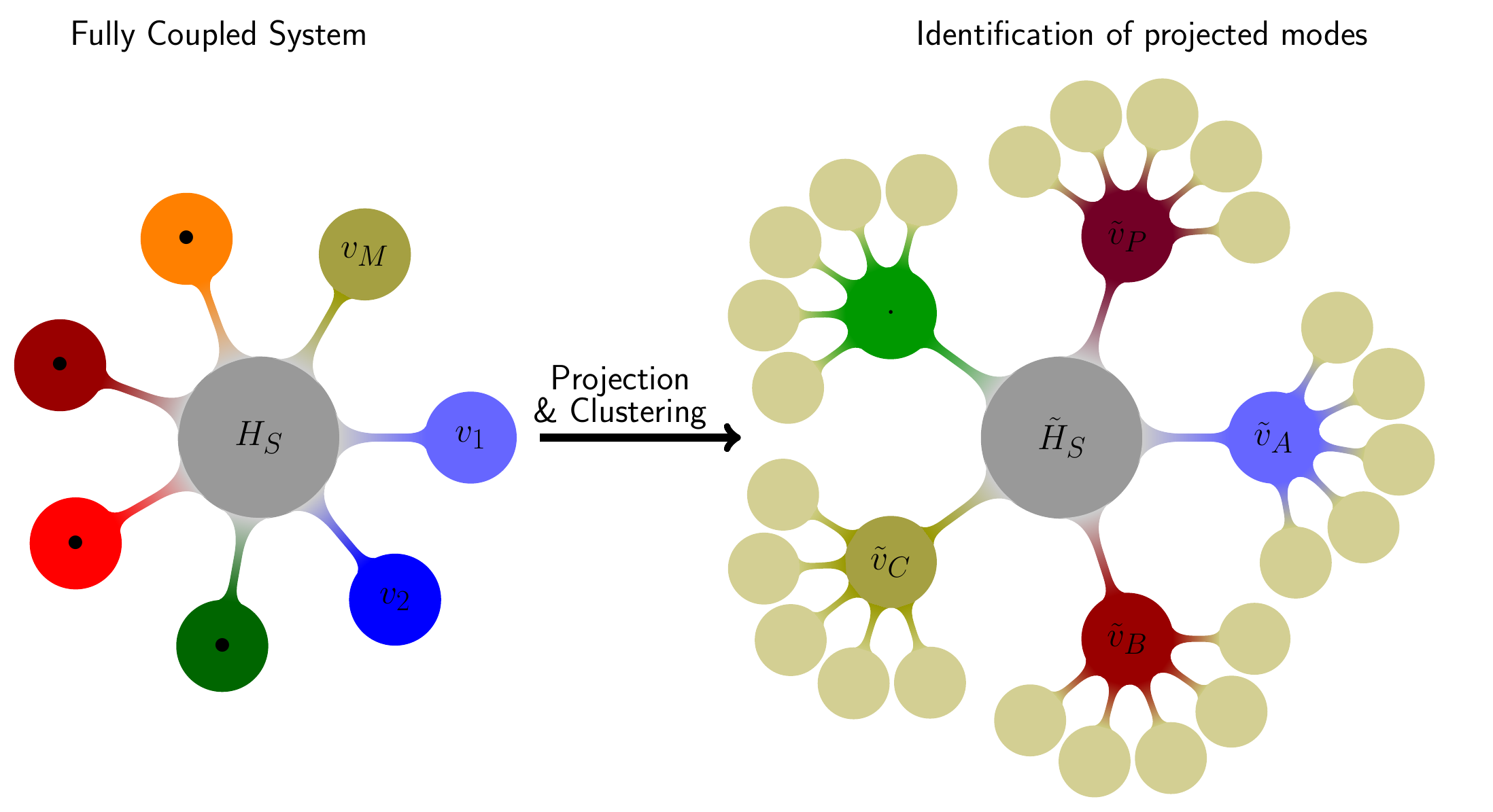}
    \caption{Schematic diagram of quantum embedding for spin-phonon interaction. The SVD projects the original phonon modes into a primary set of phonons directly coupled to central spins and a secondary phonon bath that has no coupling to the central spin but is coupled to the primary phonon bath, leading to the lifetime of the primary phonon modes.}\label{fig:1}
\end{figure}

Both SMMs and MOFs feature one or a few central spins surrounded by environments of nuclear spins and molecular phonons (or vibrations), as illustrated in Figure~\ref{fig_sph}. Typically, in transition-metal systems, the Zeeman term dominates. Disregarding the hyperfine and dipole-dipole terms, we express the spin Hamiltonian as
\begin{equation}
H_{s} = \beta{\vec{\mathbf{S}}}\cdot\bm{ g}\cdot{\vec{\mathbf{B}}}
\label{eq:spin}
\end{equation}
where $\beta = \mu_B/\hbar$ represents the ratio of the Bohr magneton to the reduced Planck constant, ${\vec{\mathbf{S}}} = (\hat{S}_x,\hat{S}_y,\hat{S}_z)$ denotes the vector spin, ${\vec{\mathbf{B}}} = (B_x, B_y, B_z)^T$ represents the magnetic field, and $\bm{ g}$ is the $g$-tensor that characterises the coupling to the external magnetic field. For \ce{VOPc(OH)8}, we computed the g-tensor as
\begin{equation}
\mathbf{g} =
\begin{pmatrix}
2.073 & -1.64 \times 10^{-2} & -9.25 \times 10^{-3} \\
-1.64 \times 10^{-2} & 2.037 & 3.50 \times 10^{-3} \\
-9.22 \times 10^{-3} & 3.58 \times 10^{-3} & 2.032
\end{pmatrix}
 \end{equation} 
using density functional theory as discussed below.
The fact that this is not diagonal poses no real problem since it can be brought into diagonal form
by rotating the inertial frame of the molecule to the external (\textit{i.e.} laboratory) frame of the applied magnetic field
giving $(g_x,g_y,g_z) = (2.081, 2.031, 2.030)$ in the rotated frame.
Although these
values differ from the experimental
values of $(g_x,g_y,g_z) = (1.966, 1.989, 1.989)$ reported in Ref.~\citenum{Ryan2021jacs} for \ce{VoPc},
we made no effort to adjust the PBE exchange/correlation functional used in our calculations
to fit our computed values
to the experimental values.


Although nuclear displacements
do not appear directly in this expression, $\bm{ g}$ depends on the local molecular geometry
via second-order perturbation theory;
\begin{align}
    g_{\alpha\beta} = g_o + \sum_e \frac{\langle \psi_g|\hat L_\alpha|\psi_e\rangle
    \langle \psi_e| \hat L_\beta|\rangle}
    {E_e-E_g},
\end{align}
where $g_o$ is the g value for a 
free electron and $\hat L_\alpha$ 
are the orbital angular momentum operators coupling the ground electronic state $|\psi_g\rangle$ to 
the unoccupied electronic states $|\psi_e\rangle$ taken at the 
equilibrium nuclear geometry.  
Of the terms entering this expression, the energy gap is most sensitive to fluctuations in
equilibrium geometry. 
Further, we can use simple symmetry arguments to deduce which orbital terms contribute to the tensor since
the integrand of each of the
matrix elements must be in a totally
symmetric irreducible representation,
i.e.
\begin{align}
    \Gamma_{A_1} = \Gamma_{g}\otimes\Gamma_{op}\otimes \Gamma_{e}
\end{align}
For a $C_{4v}$ molecule such as \ce{VOPc}, the $d_{xy}$ orbital is 
in the $B_2$ irreducible representation 
and is coupled via $\hat L_{x}$
and $\hat L_y$ the orbital angular momentum operators to the $d_{xz}$ and
$d_{yz}$ orbitals ($E$ irreducible representation)   and similarly to the $d_{x^2-y^2}$
orbital ($B_1$ irreducible representation) via the $\hat L_z$ operator 
($A_2$ irreducible representation)
Recent work by Kazmierczak {\em et al.}  indicates that the
couplings are the most sensitive 
to those vibrational normal modes that 
are also in these irreducible representations.
\cite{Ryan2021jacs}

The phonon or vibrational degrees of freedom are described by the harmonic oscillator Hamiltonian:
\begin{align}
\label{eq:phonon}
    H_b &=  \sum_{k}^{n_q} \frac{1}{2} \left(\hat{p}^2_k + \omega_k^2 \hat{x}_k^2\right)
\end{align}
where $\hat{x}_k$ and $\hat{p}_k$ represent the mass-weighted normal coordinates and their conjugate momenta, respectively, the summation runs over $n_q$ phonon degrees of freedom. 
To capture the effect of the phonons on the spin, we expand the spin Hamiltonian concerning the phonon displacements about the ground-state geometry and truncate the expansion q5 first order. The resulting Hamiltonian for the total system is of the form
\begin{align}
\label{totalH}
H &= \sum_{\alpha}^{n_s} h_\alpha \hS_\alpha + \sum_k^{n_q}\frac{1}{2}(\hat p_k^2 +\omega_k^2\hat x_k^2 ) + \sum_{\alpha k}^{n_s, n_q}g_{\alpha k}\hS_\alpha\hat x_k,
\\
H &= H_s + H_b + H_{sb}.
\label{Hsb} 
\end{align}
Here, the first term sums over the complete set of operators describing the $n_s$ spin degrees of freedom, where we have defined 
\begin{align}
    h_\alpha = \beta \sum_j^3 \bm{ g}_{\alpha j} \cdot B_j
\end{align}
as the spins and external magnetic field coupling via spin operator $\hat{S}_\alpha$.
The last term defines the coupling between the spins and the
vibrational degrees of freedom,
determined from the gradient of $H_s$ with respect to the nuclear
displacements.
\begin{align}
\label{eq:coupling}
    g_{\alpha k} = \beta \sum_j^3 \left(\frac{\partial \bm{{ g_{\alpha j}}}}{\partial \vec x_k} \right)^{(0)} B_j.
\end{align}
 The system usually has a small
number of discrete quantum levels compared to the larger degrees of freedom (DOFs) in the bath, such that $n_s \ll n_q$.

The gradient term can be computed using Density Functional Theory (DFT) calculations from software packages such as ORCA and CP2K. However, this process presents significant challenges as it requires using a finite-difference method, which is computationally demanding and must be performed for each system individually. In this work, we used ORCA (v5.0) ~\cite{orca} to compute the magnetic properties (g-tensor) of the SMMs in both equilibrium and distorted configurations. We employ the def2-TZVP base set for metal and \ce{O} elements while using the def2-SVP basis set for other lighter elements, complemented by an auxiliary def2-TZVP/C basis set for all elements. We calculate the g-tensor using the DFT method with the PBE functional.~\cite{PBE}
Initially, we optimised the molecular system's geometry using DFT and then calculated the Hessian matrix to obtain the vibrational normal modes. We then estimate the spin-phonon coupling coefficients by calculating the gradient of $g$ with respect to the nuclear coordinates using the finite difference method. Specifically, we approximate 
\begin{align}
\frac{\partial g(\boldsymbol{R})}{\partial X_{ix}}\approx \frac{g(\boldsymbol{R}+\Delta_{ix}) - g(\boldsymbol{R}-\Delta_{ix})}{2\Delta_{ix}}
\end{align}
where $X_{ix}$ denotes the $x$ Cartesian component of the $i$-th atom in the optimized geometry and $\Delta_{ix}$ is the perturbation along that direction , $\boldsymbol{R}$ represents the equilibrium grometry. To achieve this, we distort the optimised geometry with small forward and backward displacements ($\pm 0.01$ \AA) and compute the $g$-tensors of the distorted geometries.

For a molecule with $N$ atoms, there are $3N-6$ normal modes. However, not every vibrational mode will contribute equally to the
coupling.  
 To understand the effectiveness of each phonon displacement towards spin-phonon coupling,  we computed the $l_2$ norms of the resulting g-tensor gradient with respect to dimensionless displacements $q_k$ for \ce{VOPc(OH)8} via
 \begin{equation}
     || \partial g/\partial q_k||^2 = \sum_{i,j}^3 \sqrt{\left(\partial g_{ij}/\partial q_k\right)^2}.
 \end{equation}
Consequently, just as the $l_2$ norm of a vector gives its length, the $l_2$ norm of the gradient tensor gives its overall amplitude and indicates the participation of each normal mode $k$ in the spin-phonon coupling. Although spin-phonon coupling linearly depends on the applied magnetic field, the gradient still determines the participation of each individual phonon; therefore, using the $g$-tensor gradient instead of the full spin-phonon coupling highlights the contribution of each phonon in the spin-phonon coupling.
 
\begin{figure}[htb]
 \includegraphics[width=0.5\textwidth]{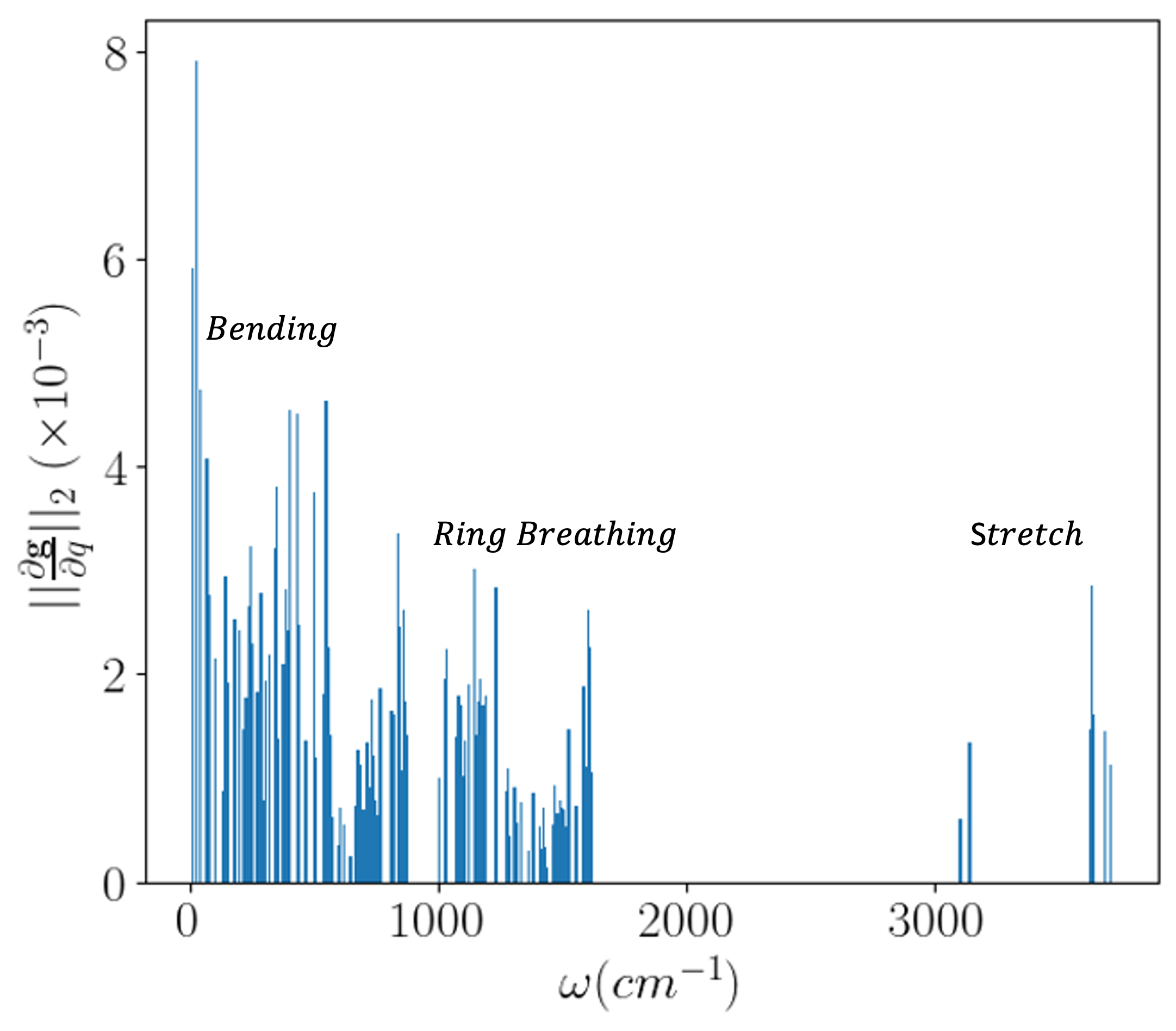}
    \caption{$l_2$ norm of the g-tensor derivative ($\vec\nabla  g$) with respect to each phonon mode for 
    \ce{VOPc(OH)8} as obtained
    from DFT calculations. 
    The normal mode frequencies ($cm^{-1}$) are on the x-axis, with $l_2$ norms on the y-axis (unitless). The histograms show a distinct band structure, with the first group of frequencies representing the bending modes of
    the 
    The second group, the phenylocyanine ring, belongs to the ring-breathing modes, and the highest-frequency modes are the OH and CH stretch modes.
    }
    \label{fig: histogram}

\end{figure}

An analysis of Figure~\ref{fig: histogram} reveals that the few lowest frequency modes have the highest norm of the $g$-tensor gradient, although the high-frequency modes also play a role. Furthermore, the amplitude of the norm $l_2$ for all modes is of the order of $10^{-3}$ (compared to the values of the g-tensors of $\approx 2$). Therefore, spin-phonon coupling is relatively weak
in \ce{VOPc(OH)8} MOFs, which explains their relatively long $T_1$ 
relaxation times and their candidacy for molecular qubits.

Defining the entire system/bath Hamiltonian is straightforward, but solving its complete quantum dynamics is a significant challenge for any reasonably sized system. This difficulty arises because quantum dynamics exhibits superlinear scaling with the bath DOF 
 which scales as $3N$, where $N$ is the number of atoms involved. This challenge can be addressed by implementing a series of systematic projections, as illustrated in Fig.~\ref{fig:1}. These projections simplify the fully coupled (star) system into a series of independent branches, where only a few bath degrees of freedom are directly coupled to the system, while the rest act as a thermal reservoir. This approach effectively transforms a high-dimensional and intractable quantum problem into a more manageable one that can be solved using contemporary quantum dynamics approaches.

 \subsection{Decomposition}

The general model for a quantum system that interacts with an environment can be represented as a tensor product of system operators ($\hS$) and bath ($\hB$) operators, with coupling constants $g$. The Hamiltonian of this system, as shown in Eq.~\ref{totalH}, includes the summation over all bath degrees of freedom (DOFs) in the interaction term $H_{sb}$. As previously mentioned, Fig.~\ref{fig:1}  graphically presents this situation in its left diagram. The objective is to reduce the number of DOFs that directly interact with the system through projective transformations. These transformations lead to the creation of two sets of new DOFs. One set, called \emph{system-phonons}, interacts directly with the spin system, and the other, called \emph{bath-phonons}, interacts only with the first set, as depicted in the right panel of Fig.~\ref{fig:1}. To achieve this, Bittner\textit{et al.} introduced a Mode Projection scheme,
which determines the optimal partitioning
of the entire dynamical system into interacting subcomponents
~\cite{modeprojection2009jcp, projection2014jpca}.
The approach has been used successfully to determine electronic charge and energy
transfer in small molecular systems
with various degrees of complexity\cite{Yang:2017aa,Yang:2018ab}
using \textit{ab initio} derived electron/phonon couplings.

This work presents a single-value decomposition (SVD)--based mode projection method for reducing the spin-phonon coupling's dimension.
Consider the coupling matrix and its SVD, defined as
\begin{equation}
    \bm{g} = [g_{\alpha k}]_{n_s\times n_q} = \bm U_{n_s \times n_s} \cdot \bm \Sigma_{n_s\times n_q} \cdot \bm V^\dagger_{n_q \times n_q}
\end{equation}
where $\Sigma$ is a matrix of singular values with at most $\min(n_s, n_q)$ nonzero singular values. Taking column vectors of $V^\dagger$ corresponding to non-zero singular values and creating projection operator
\begin{equation}
\mathbb{P} =  \sum_r |V_r^\dagger \rangle \langle  V_r^\dagger|,
\end{equation}
 projects the phonon modes into reduced dimensional subspace. Likewise its complement
\begin{equation}
\mathbb{Q} = \mathbb{I}-\mathbb{P},
\label{eq:14}
\end{equation}
projects them into the complement subspace. Using these projection operators, one could reduce the dimensionality of the spin-phonon coupling; however, to convert this projection scheme into a canonical transformation, we proceed with diagonalisation of the Hessian in the two projected sub-spaces. In the original normal mode basis, the Hessian is a diagonal matrix defined by $\Omega^2 = [\omega_{ii}^2]$. Using the resolution of identity  as a sum of projection operators,
\begin{align}
    \Omega^2  &= \left(\mathbb{P}+\mathbb{Q}\right)\Omega^2 \left(\mathbb{P}+\mathbb{Q}\right)\\ \nonumber
    &= \mathbb{P} \Omega^2 \mathbb{P} + \mathbb{Q} \Omega^2 \mathbb{Q} + \left( \mathbb{P} \Omega^2 \mathbb{Q} +  \mathbb{P} \Omega^2 \mathbb{Q} \right)\\ \nonumber
    &= \Omega^2_{PP} + \Omega^2_{QQ} + \Omega^2_{PQ} +\Omega^2_{QP}
\end{align}
\begin{widetext}
Diagonalising $\Omega^2_{PP}$ and $ \Omega^2_{QQ} $ provides the frequencies $\omega_k^2$ for the new system and bath modes, respectively. Since the number of DOF of phonons is much larger than the spin DOF, $\Omega^2_{PP}$ will have at most $n_s$ nonzero eigenvalues, while $\Omega^2_{QQ}$ will have $n_q-n_s$ nonzero eigenvalues. The corresponding eigenvectors or nonsingular eigenvalues for $\Omega^2_{PP} (\{\tilde x^P\})$ and $ \Omega^2_{QQ} (\{\tilde x^Q\})$ constitute the system and bath modes, respectively. And $\Omega^2_{PQ}$ defines the couplings between the system and the bath modes.
Using projected modes, the phonon and interaction terms of the total Hamiltonian Eq.~\ref{totalH} become
\begin{align}
H_{b} + H_{sb} 
&= 
\sum_{\alpha r}\tilde g_{\alpha r}S_\alpha\tilde x_r
+
\sum_r^{n_s}\frac{1}{2}(\tilde p_r^2 +\omega_r^2\tilde x_r^2 ) 
+ \sum_j^{n_q-n_s}\frac{1}{2}(\tilde \pi_j^2 +\omega_j^2\tilde y_j^2 ) 
 + \sum_{r,j}^{n_s,n_q-n_s} \gamma_{r,j} \tilde x_r \tilde y_j.
\end{align} 
The first term reflects the
{\em direct} coupling between the
$n_s$ spin degrees of freedom and
$n_r$ projected modes $\{\tilde x_r\}$ with momenta $\{\tilde p_r\}$ (note, $n_r\leq n_s$).
These are the {\em primary} modes since they are directly coupled to the spin variables via projected forces $\tilde g_{\alpha r}$.  The second term represents the harmonic oscillator Hamiltonian for these primary modes. The third term is a harmonic oscillator representation of the {\em residual} modes $\{\tilde y_j\}$ with conjugate momenta $\{\tilde\pi_j\}$. 
The residual modes are linearly coupled to the primary modes via $\gamma_{r,j}$ in the fourth term. 
\begin{figure*}
    \centering
    \includegraphics[width=0.3\columnwidth]{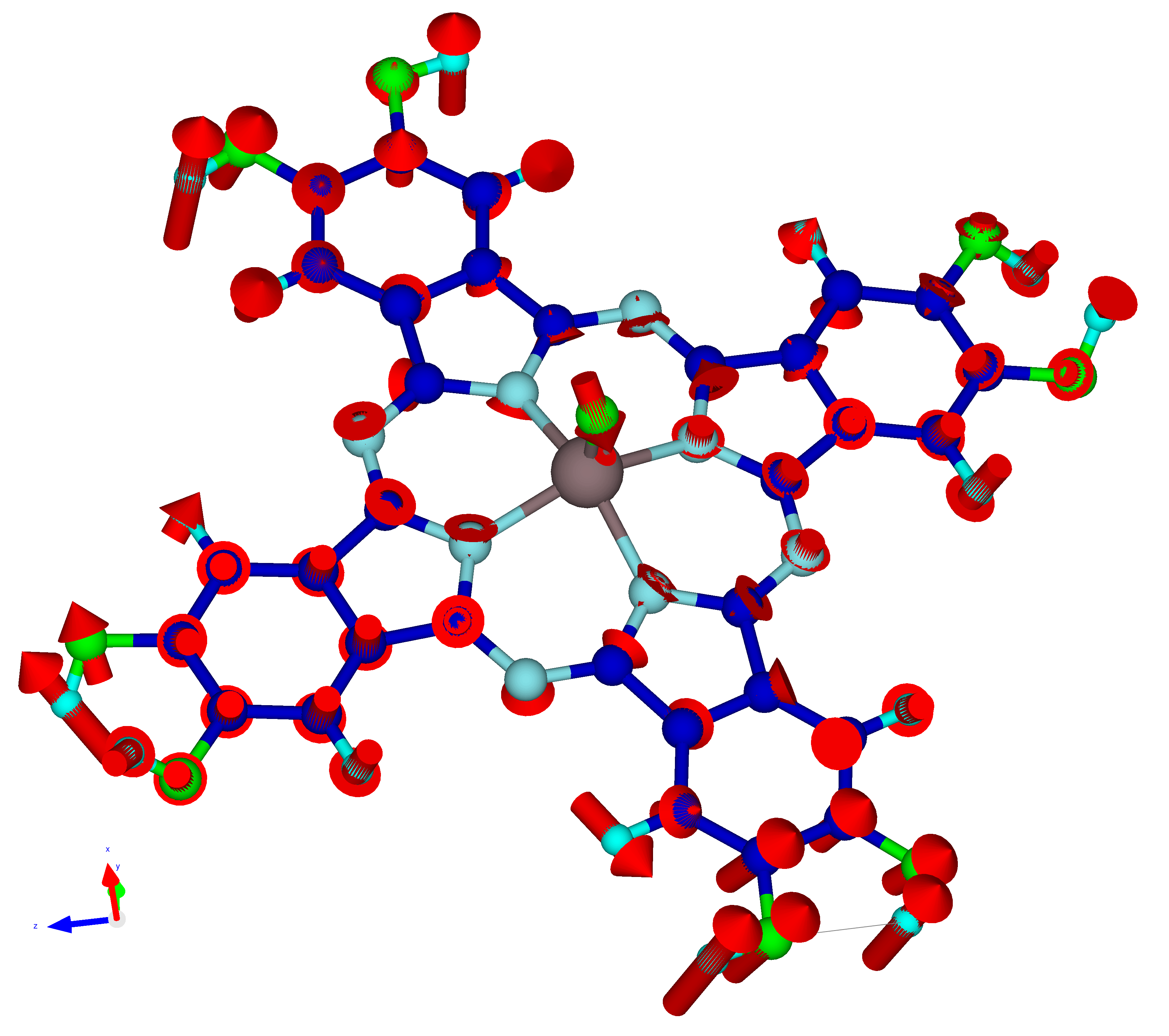}
    \includegraphics[width=0.3\columnwidth]{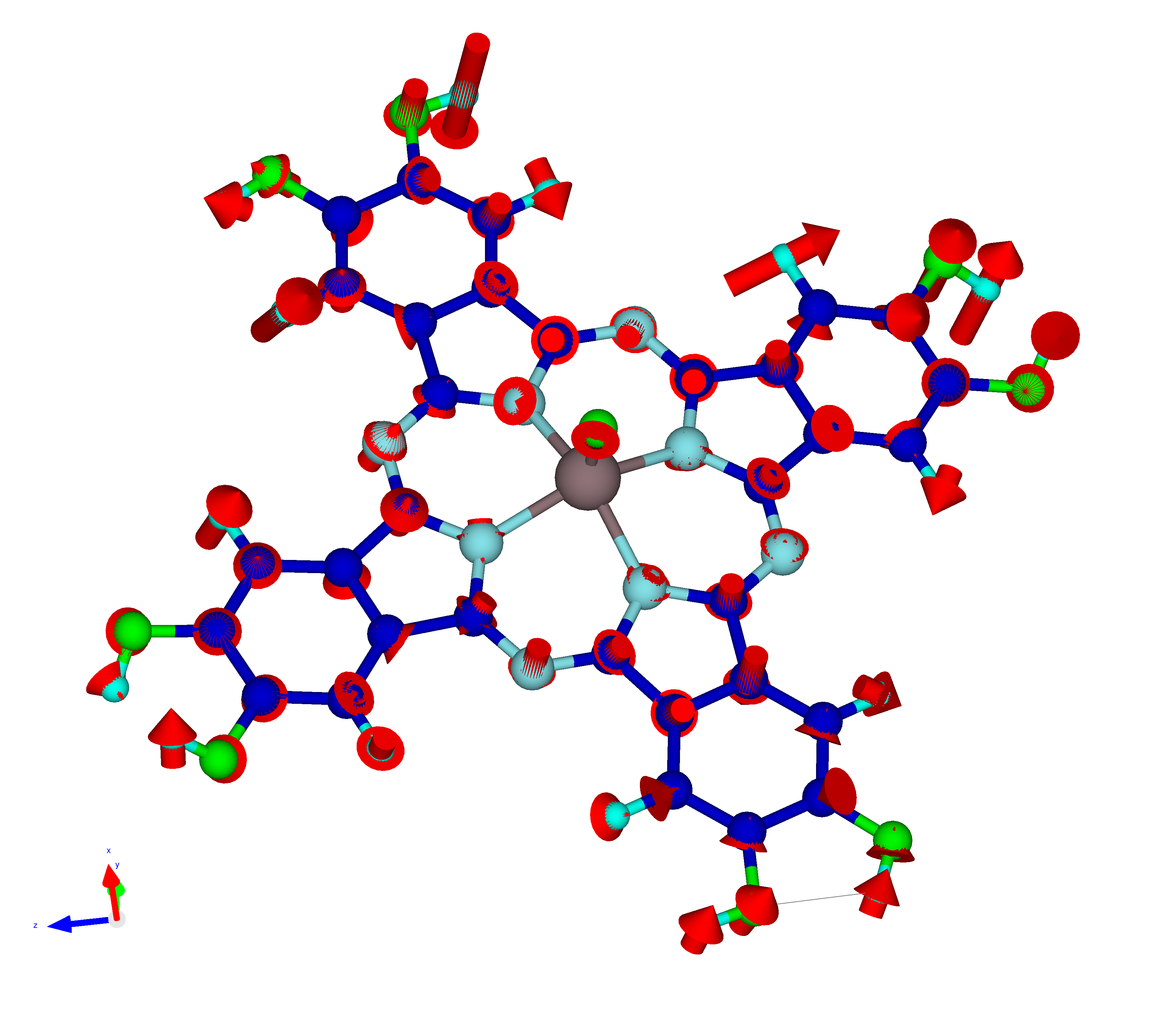}
    \includegraphics[width=0.3\columnwidth]{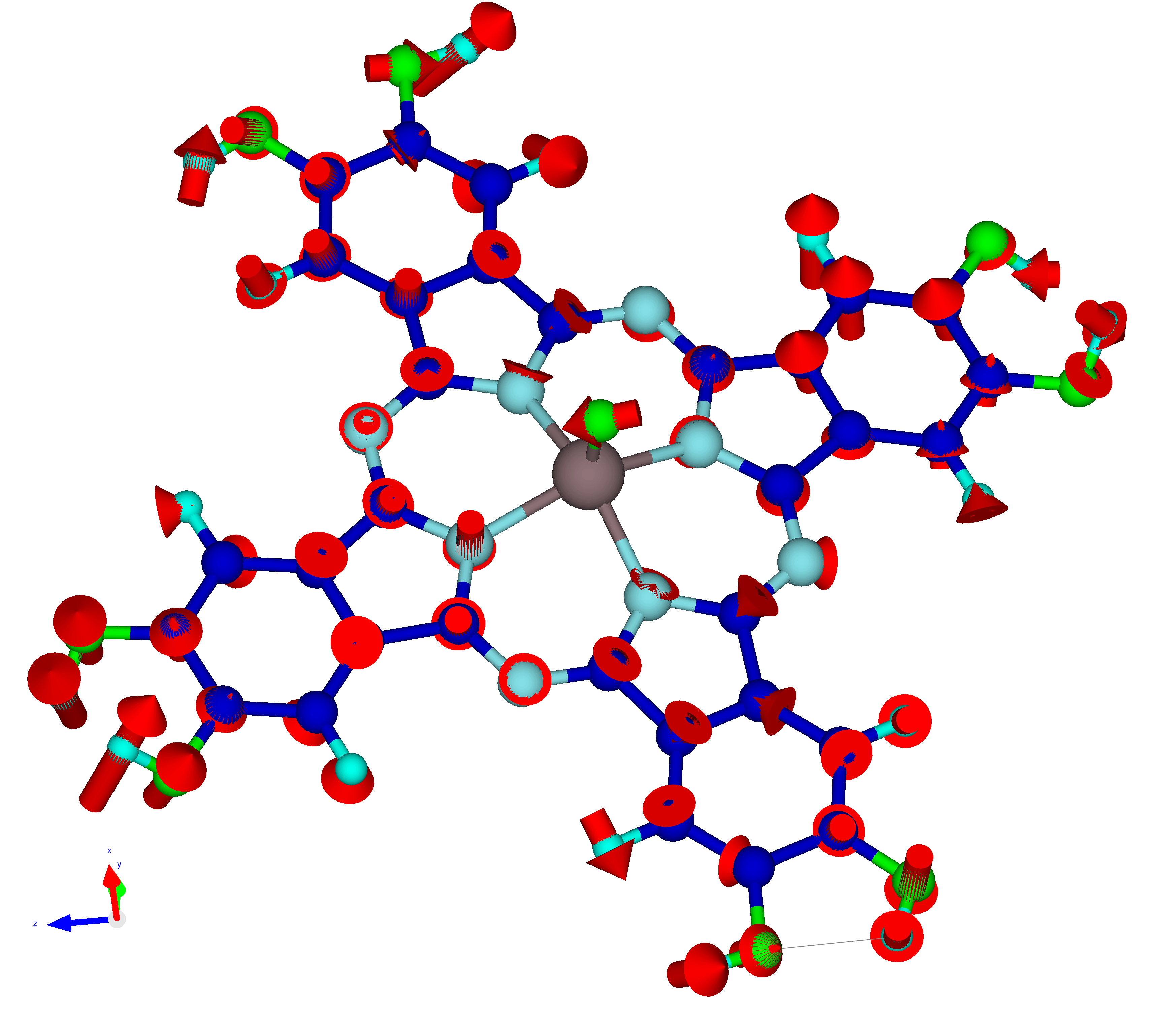}
    \includegraphics[width=0.6\textwidth]{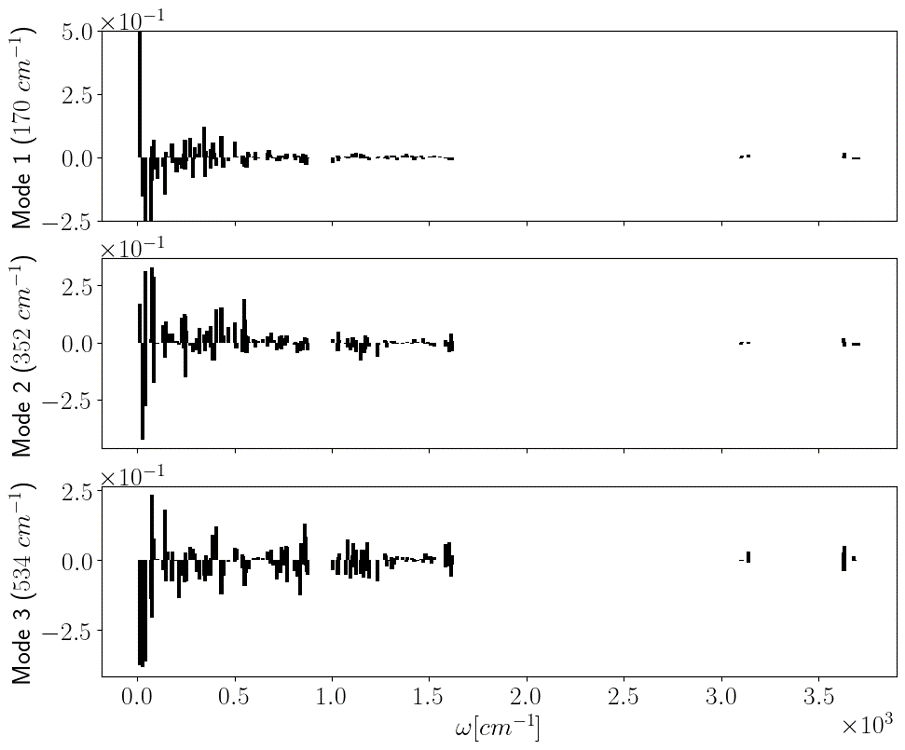}
    \caption{Projected modes are linear combinations of original phonons. The contribution of initial phonons in projected system modes 1, 2, and 3 is given in the top, middle, and bottom panels, respectively.}
    \label{fig:projected_mode_components}
\end{figure*}
\end{widetext}
\section{Results}

Using this SVD mode projection approach and the computed spin-phonon coupling matrix for \ce{VOPc(OH)8)}, we determined three primary modes at frequencies 170.05, 352.20, and 533.980 $cm^{-1}$, respectively.
These projected modes are linear combinations of the normal vibrational 
modes of the molecule. 
The atomic displacements within the \ce{VOPc(OH)8} molecule associated with the primary modes are shown in Fig.~\ref{fig:projected_mode_components} (top). The primary mode displacements include the bending and twisting of the phthalocyanine structure and the movement of the central oxygen. The couplings of these modes with spin through spin operators $\hat S_x$, $\hat S_y$, and $\hat S_z$ are given in the Supplemental Information.
Fig.~\ref{fig:projected_mode_components} (bottom) shows the contribution of each original phonon to projected modes 1, 2, and 3, at frequencies 170.05, 352.20, and 533.980 $cm^{-1}$, respectively. In particular, original modes with low frequencies are dominant contributors, but even high-frequency modes make small contributions to the projected modes.
 However, it should be pointed out that replacing the terminal OH groups with methyl groups affects only the very high frequencies assigned to the OH stretch. They are replaced by the CH stretch modes. Because the spectral densities drop off quickly with phonon frequency, replacing the OH group with the methyl group does not affect the spin dynamics. This observation is supported by identical spin-phonon couplings and identical phonon frequencies of the central molecular structure for the two cases. 

 A useful measure of the decomposition efficiency can be obtained by calculating the von Neumann entropy. 
 \begin{align}
\label{eq:entropy}
     S = -\sum_n r^2_n \ln r^2_n
 \end{align}
 where $r_n = \lambda_n/||\lambda_n||
 $ 
 are the components of the vectors representing phonons. 
Using Eq.~\ref{eq:entropy} and considering $r_n$ as elements of the projected vectors, the von Neumann entropies for these vectors are $1.96$, $3.22$, and $3.42$ for the first, second, and third projected modes, respectively. Unlike the original normal modes,  the projected modes exhibit an increased entropy, highlighting the efficiency obtained by compressing the
information carried by
the full environment into
a small number of collective
degrees of freedom.
 In particular, the number of primary vectors that SVD returns is proportional to the matrix's rank, which is at most 3 for a single spin-1/2 system.
The original normal coordinates
can be recovered by diagonalizing the last three terms.
As a result of this transformation, the coupling between spin and phonons involves a sum over only the system modes; since $n_s \ll n_q$ is generally satisfied in molecular magnets, the SVD-based projection can significantly reduce the number of phonon modes in the simulations.

\begin{figure*}[htb]
    \centering
    \subfigure[]{\includegraphics[width=0.31\textwidth]{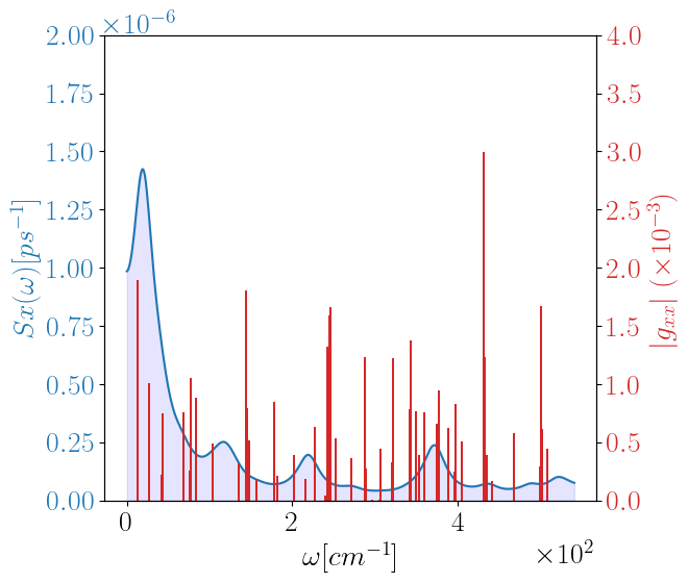}}
    \subfigure[]{\includegraphics[width=0.32\textwidth]{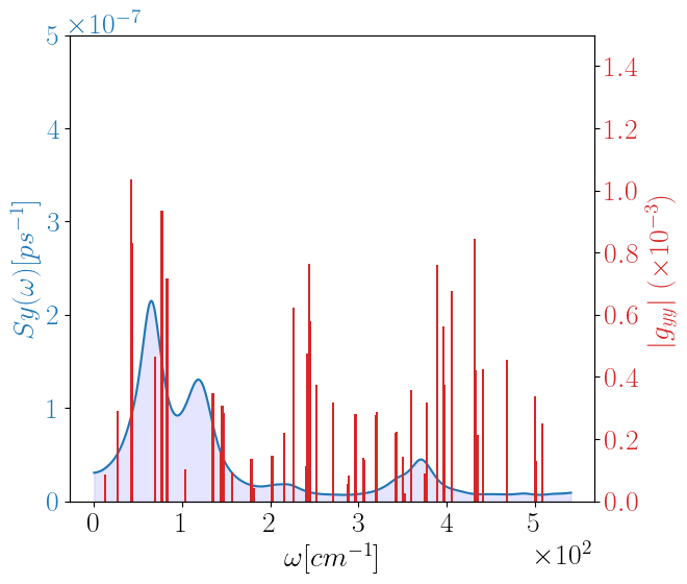}}
    \subfigure[]{\includegraphics[width=0.33\textwidth]{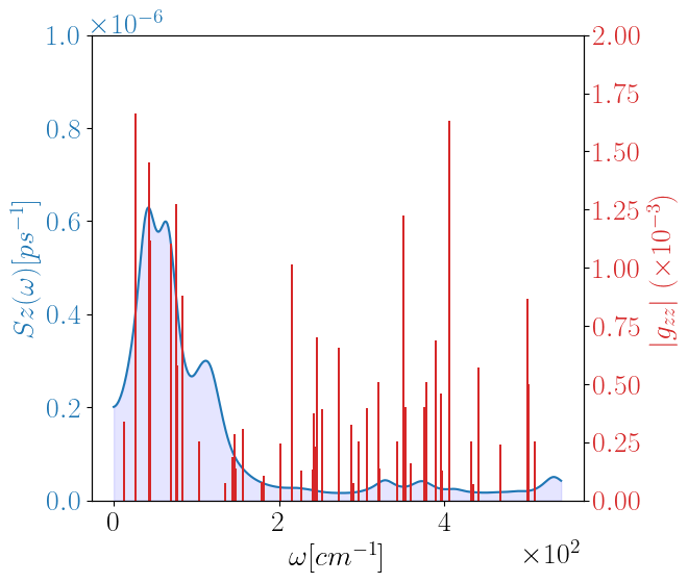}}
    \caption{Spectral density (Eq.~\ref{spectraldensity}) of the phonons specific to $\sigma_x$ (a), $\sigma_y$(b), and $\sigma_z$(c)  (blue curve in the bottom, middle, and top panels, respectively) at 300 K temperature with a magnetic field of 1 T along the molecular $z$-axis. The derivatives of the g-tensor elements are shown in red. The spectral density for high-frequency modes is low even at ambient temperature. Ultimately, this spectral density affects phonons' participation in the spin system's Redfield dynamics, penalising the higher frequency components.  
   Movies of the nuclear motions are given in the Supplemental Information.}
    \label{fig:parallel_hist}
\end{figure*}

The couplings, both for the full set of modes and projected modes, contribute to the bath spectral densities required by the Redfield master equation for simulating the quantum dynamics of the spin system. 
Expressions for these spectral densities are given in the Supplemental Information
and take the usual form of a sum over all bath modes. 
Fig.~\ref{fig:parallel_hist} presents the spectral density for each of the three spin components without applying SVD and using full set of original normal modes. Additionally, the g-tensor gradients projected along each normal mode are also given. The spectral densities decay rapidly with increasing phonon frequency, indicating that low-energy phonons predominantly contribute to the bath spectral densities in the Redfield master equation. Furthermore, the $x$ and $z$ components dominate the spin/phonon spectral densities, whereas the components of the g-tensor gradient are almost uniformly distributed throughout the vibrational spectrum.

\begin{figure*}[htb]
    \centering
    \includegraphics[width=1\textwidth]{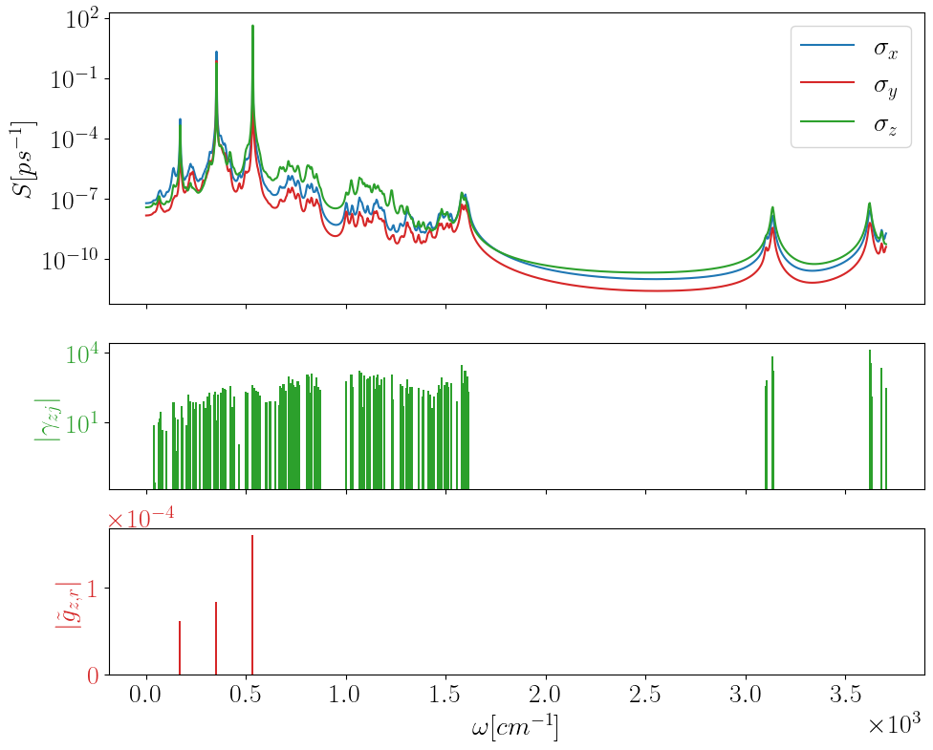}
    \caption{Spectral density for mode projection at 300K and 1T (top), couplings between system modes and bath modes (middle), and couplings between spin and system modes (bottom).
 }
\label{fig:spectral_densities_system_and_bath_modes}
\end{figure*}

After SVD mode projection, we compute the spectral density of the primary modes using the couplings between the spin and primary modes and the phonon-phonon couplings between the primary and residual modes. Fig.~\ref{fig:spectral_densities_system_and_bath_modes} shows the spectral density along with the spin-phonon and phonon-phonon couplings. For clarity, we show the spectral density and spin-phonon couplings through the spin operators $\hat S_x, \hat S_y$, and $\hat S_z$. The spectral density exhibits a unique lineshape, indicating resonances at both the primary and residual mode frequencies. This characteristic results from the driven-dissipative harmonic oscillator nature of the primary modes, as detailed in Eq.~\ref{spectraldensity} and the Supplementary Information that accompanies this paper. 

\subsection{Spin dynamics with original modes}
With the spin-phonon couplings at hand, 
we can investigate the spin relaxation dynamics of vanadyl phthalocyanine (\ce{VOPc(OH)8}) using the Bloch-Redfield
equations of motion for 
the reduced density matrix of the isolated spin,
\begin{align}
    \partial_t \rho_s = -i[H_s,\rho_s] 
    + {\cal R}(\rho_s)
\end{align}
where $({\cal R}(\rho_s)$ is the
Redfield superoperator 
\begin{align}
    ({\cal R}(\rho_s))_{\alpha\beta}
    = \sum_{\gamma\delta}
    R_{\alpha\beta;\gamma\delta}[\rho_s]_{\gamma\delta}
\end{align}
which 
depends upon the spectral density
of the bath as implemented in the QuTip (v.5) 
quantum dynamics package.~\cite{JOHANSSON20121760,JOHANSSON20131234}
This is given explicitly by 
\begin{equation}
\label{spectraldensity}
    S_\alpha(\omega) = \pi \hbar \sum_k^{n_q} \frac{g_{\alpha k}^2}{\omega_k}\left[\left(n(\omega_k)+1\right) \delta(\omega-\omega_k) + n(\omega_k)\delta(\omega+\omega_k)\right]
\end{equation}
where $n(\omega)$ are the thermal phonon occupations given by Bose-Einstein statistics.
For computational purposes, we replace the delta $\delta(\omega\pm\omega_j)$ functions with the Gaussian functions \textit{i.e.}. $\frac{1}{\sqrt{2\pi}\Delta\omega} e^{-\frac{\left(\omega\pm\omega_j\right)^2}{2\Delta\omega}}$
using $\Delta\omega =2 \text{ meV} \approx 16 \ cm^{-1}$ as per the FTIR spectra of VOPc(OH) and VOPc(Me) compounds. The spectral densities and g-tensor derivatives for the normal modes of single molecule VOPc(OH)8 are shown in Fig.~\ref{fig:parallel_hist}

\begin{figure*}[htb]
\subfigure[]{\includegraphics[width=0.3\textwidth]{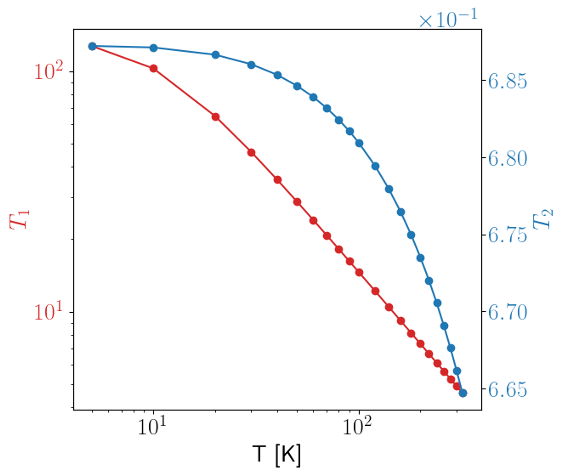}}
\subfigure[]{\includegraphics[width=0.3\textwidth]{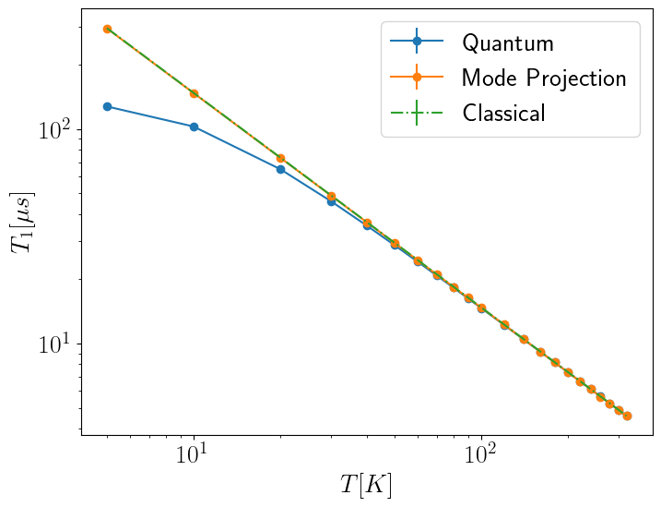}}
\subfigure[]{\includegraphics[width=0.3\textwidth]{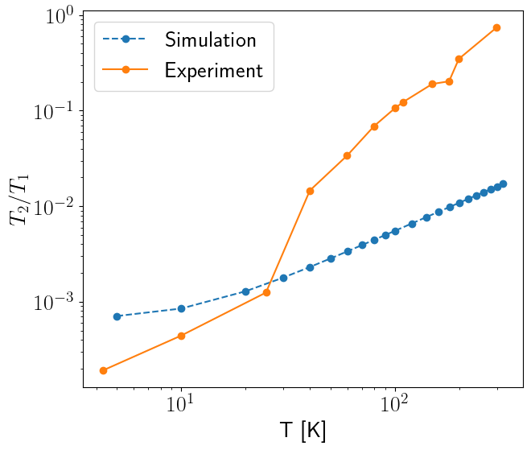}}
    \caption{(a) Temperature dependence of $T_1$ and $T_2$ (red and blue, respectively) calculated from Redfield simulations using the full set of modes at 1 Tesla Magnetic field.
    (b) $T_1$ as a function of temperature using a complete set of modes with quantum and classical spectral density and projected modes with classical spectral density. The green curve overlays the orange curve.
    (c) Temperature dependence of the ratio between $T_1$ and $T_2$ calculated from Redfield simulations (blue-dashed) and from experimental observations (yellow-solid) reported by Atzori {\em et al.} at 345 mT.~\cite{Atzori2016} }
  \label{fig:redfield_combined}

\end{figure*}

Fig.~\ref{fig:redfield_combined} (a) shows the temperature dependence of the $T_1$ population relaxation time and the $T_2$ pure dephasing
time produced by our model.  As expected,  $T_1$ shows a marked dependence on the system's temperature, whereas $T_2$ time is nearly constant over the same range due to its dependence on nuclear hyperfine interaction that is ignored in the current approximation.
Similarly, Fig.~\ref{fig:redfield_combined} (b) 
compares $T_1$ using the classical versus quantum
versions of the spectral
density functions. 
As expected, the two
diverge at low temperatures. 
These results provide
a baseline for comparison
against any error
introduced by the mode projection
method.  
In fact, over the entire temperature range,
the mode-projection/reduction approach gives identical
results compared to the standard Redfield
treatment as shown in Fig.~\ref{fig:redfield_combined} (b) and discussed in the next section.

Finally, we compare
our {\em ab initio} model with recent experimental data on the
VoPC system reported by Atzori {\em et al.} 
taken at 345 mT.
In Fig.~\ref{fig:redfield_combined} (c), we compare the ratio of $T_2/T_1$.  
Analytically, 
this ratio should equal 2 if the spectral density takes a simple Lorenzian form. 
In general, the comparison
between the model and the experimental data gives
The correct overall trend is at low temperatures but differs by several orders of magnitude at higher temperatures. This can be accounted for because the simulations only account for a single molecule in a vacuum, thus neglecting the effects of low-energy phonons in solid-state systems. Furthermore, the linear dependence on the phonon displacement encoded in the spin-phonon interaction Hamiltonian does not account for
higher-order inelastic processes that become the
dominant mechanism at low magnetic fields and intermediate temperatures compared to direct phonon coupling.~\cite{Lunghi_spin_raman}

\subsection{Spin Dynamics with Projected Modes}

Having defined 
the projected modes, 
we can perform \textit{numerically
exact} quantum simulations
within the tensor space
defined by the central 
spin and the few projected 
modes using the
reduced Hamiltonian
\begin{align}
    H_{red}
= \sum_\alpha h_\alpha \hat S_\alpha + \sum_{\alpha r}
\tilde g_{\alpha,r}S_\alpha
\tilde x_r + \sum_{r}^{n_s}
\frac{1}{2}(\tilde p_r^2 + \omega_r^2 x_r^2)
\end{align}
and treating the $n_q$ residual modes as a thermal 
bath for the 
primary ($\tilde x_r$) 
degrees of freedom.

Using this, we evaluated the effectiveness of projected modes in capturing the influence of the phonon bath on the spin system by simulating the spin dynamics over the lifetime $T_1$. This simulation employs the spectral density for the primary modes within the framework of the Redfield master equation, as previously discussed. Fig.~\ref{fig:redfield_combined} shows the resulting temperature dependence of $T_1$. Since we compute the spectral density of the projected modes using a classical formalism, the $T_1$ results using projected modes match those obtained using the full set of phonons classically. This complete match indicates that the projected modes successfully capture the phonon bath originally represented by 192 phonons.
Additionally, in the Supplementary Information, we compare the performance of projected modes against a selection of original phonon modes using a trivial cutoff on spin-phonon coupling by applying a threshold at $35\%$ of the maximum spin-phonon coupling. 
In doing so, we can reproduce the
relaxation $T_1$ as a function of temperature, indicating that the SVD approach
is highly effective in identifying
an optimal partitioning of the phonon
environment.

\section{Summary}
\label{sec:summary}

This study presents an innovative quantum embedding approach for studying spin-phonon dynamics in molecular magnets. The proposed method effectively investigates the spin-lattice relaxation process with high precision. We employed this approach to analyse the spin dynamics in single molecular magnets. Through this methodology, we have gained a more profound understanding of the collective intra-molecular motions 
governing spin relaxation in molecular magnets, which is critical for optimising their performance in various applications. Moreover, this research establishes a foundation for rationalising new materials with tailored magnetic properties suitable for quantum information processing and other technological applications.


In addition to shedding light on the collective intramolecular motions that dominate spin-phonon coupling, our embedding approach enables the quantum treatment of phonons when their energies are comparable to spin splitting. Such scenarios can occur in the high magnetic field limit and in solid-state systems that feature low-frequency crystal phonons with energies similar to the Zeeman splitting of the spin system.
In our accompanying paper\cite{paper2}, 
we explore the quantum entanglement between electronic spin and projected modes.  Here, the projected modes serve as conduits
for channelling quantum information from the
spin to the remaining or ``residual'' degrees
of freedom corresponding to the modes spanning the $\mathbb{Q}$ subspace from Eq.~\ref{eq:14}.

In our future work, we intend to expand the quantum embedding method to include spin-spin interactions in molecular magnets. This extension aims to provide a deeper understanding of the interplay between spin-spin and spin-phonon interactions, elucidating their roles in shaping the overall relaxation dynamics of molecular qubits. By exploring spin-spin interactions, we anticipate gaining enhanced insights into these complex systems, which will contribute to developing advanced materials for quantum computing and other magnetism applications. This proposed expansion is expected to significantly broaden the scope and utility of the quantum embedding approach, thereby augmenting its value as a research tool within the domain of molecular magnets.

\vskip6pt

\enlargethispage{20pt}

\begin{acknowledgments}
The research presented in this article was supported by the LANL LDRD program (number 20220047DR). 
LANL is operated by Triad National Security, LLC, for the National Nuclear Security Administration of the U.S. Department of Energy (contract no. 89233218CNA000001). The authors thank the LANL Institutional Computing (IC) programme for access to HPC resources. 
The work at the University of Houston was funded in part by the National Science Foundation (CHE-2404788)  
and the Robert A. Welch Foundation (E-1337).
This work was performed, in part, at the Center for Integrated Nanotechnologies, an Office of Science User Facility operated for the U.S. Department of Energy (DOE) Office of Science by Los Alamos National Laboratory (Contract 89233218CNA000001) and Sandia National Laboratories (Contract DE-NA-0003525).
\end{acknowledgments}
\section*{Data Availability Statement}
The data supporting this study's findings are available from the corresponding author upon reasonable request. 

\section*{Author Contribution Statement}

    The authors confirm their contribution to the paper: study conception and design: ERB, YZ, and AP; data collection and simulations: NY; analysis and interpretation of results: NY, YZ, AP, ERB; draft manuscript preparation: NY, ERB. All authors reviewed the results and approved the final version of the manuscript.


\vskip2pc

\bibliographystyle{RS}
\bibliography{CollectedReferences}

\end{document}